\documentclass[aip, amsmath, amssymb, reprint,
]{revtex4-1}
\usepackage{graphicx}
\usepackage{dcolumn}
\usepackage{bm}
\usepackage[mathlines]{lineno}
\usepackage[utf8]{inputenc}
\usepackage[T1]{fontenc}
\usepackage{mathptmx}
\usepackage{etoolbox}

\makeatletter
\def\@email#1#2{%
 \endgroup
 \patchcmd{\titleblock@produce}
  {\frontmatter@RRAPformat}
  {\frontmatter@RRAPformat{\produce@RRAP{*#1\href{mailto:#2}
  {#2}}}\frontmatter@RRAPformat}
  {}{}
}%
\makeatother

\begin{document}


\title{Efficient Spintronic THz Emitters Without External Magnetic Field}
\author{Amir Khan}
\affiliation{New Materials Electronics Group, Technical University of Darmstadt, Merckstr. 25, 64283 Darmstadt, Germany}
\email{amir.khan@tu-darmstadt.de}
\author{Nicolas Sylvester Beermann}
\affiliation{Faculty of Physics, Bielefeld University, Universitätsstraße 25, 33615 Bielefeld, Germany}
\author{Shalini Sharma}
\affiliation{New Materials Electronics Group, Technical University of Darmstadt, Merckstr. 25, 64283 Darmstadt, Germany}
\author{Tiago de Oliveira Schneider}
\affiliation{New Materials Electronics Group, Technical University of Darmstadt, Merckstr. 25, 64283 Darmstadt, Germany}
\author{Wentao Zhang}
\affiliation{Faculty of Physics, Bielefeld University, Universitätsstraße 25, 33615 Bielefeld, Germany}
\author{Dmitry Turchinovich}
\affiliation{Faculty of Physics, Bielefeld University, Universitätsstraße 25, 33615 Bielefeld, Germany}
\author{Markus Meinert}
\affiliation{New Materials Electronics Group, Technical University of Darmstadt, Merckstr. 25, 64283 Darmstadt, Germany}
\email{markus.meinert@tu-darmstadt.de}

\date {\today}

\begin{abstract}
We investigate the performance of state-of-the-art spintronic THz emitters (W or Ta)/CoFeB/Pt with non-magnetic underlayer deposited using oblique angle deposition. The THz emission amplitude in the presence or absence of an external magnetic field remains the same and remarkably stable over time. This stability is attributed to the enhanced uniaxial magnetic anisotropy in the ferromagnetic layer, achieved by oblique angle deposition of the underlying non-magnetic layer. Our findings could be used for the development of practical field-free emitters of linearly polarized THz radiation, potentially enabling novel applications in future THz technologies.
\end{abstract}
\maketitle

The spin of the electrons in conjunction with their charge has recently\cite{Kampfrath2013} paved the way to generating the terahertz (THz) radiation that covers the range of 0.1-30\,THz using the devices called spintronic THz emitters (STEs). These are heterostructures of ferromagnetic (FM)/non-magnetic (NM) thin films with the magnetization in the film plane. The total metal thicknesses are ideally around 4 to 6 nanometers \cite{Seifert2016}. An ultrashort laser pulse excites the electrons and drives a spin current from the FM layer to the NM layer(s), which is converted into an in-plane charge current via the inverse spin Hall effect (ISHE)\cite{Jiao2013,Sinova2015} in the NM layer(s). This short charge current burst then radiates a single cycle THz pulse\cite{Seifert2017}. The amplitude and polarization of the THz wave are independent of both the pump wavelength and polarization\cite{Kumar2023,Kampfrath2013}. Notably, this process is also responsible for the demagnetization of the FM layer in femtosecond timescale (ultrafast laser-driven demagnetization\cite{Rouzegar2022}).
\begin{figure*}
\includegraphics[height = 90mm]{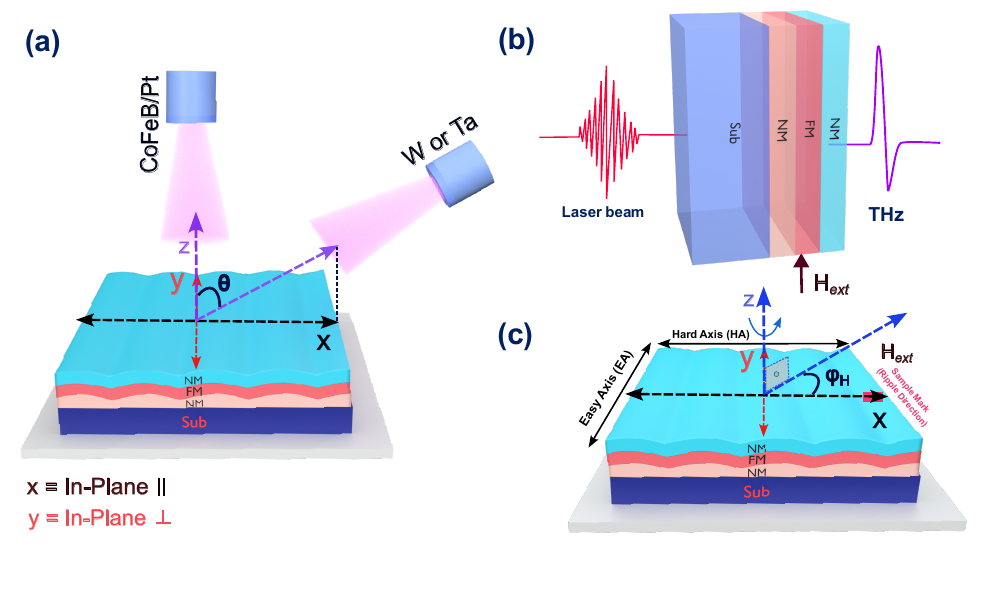}
 \caption{\label{fig : Description} Spintronic THz emitters deposition and measurement geometry. (a) W or Ta seed layer deposited at $\theta$ = 0\textdegree{}, 30\textdegree{} and 60\textdegree{}. (b) A typical NM/FM/NM STEs geometry. (c) The THz emission measurement geometry: The sample is placed in the x-y plane with the laser beam propogating along the z-axis and $\varphi_{H}$ denotes the azimuthal angle between sample mark and in-plane applied external magnetic field, where the sample mark represents incoming material flux direction or ripple direction for obliquely deposited NM underlayer.}
\end{figure*}

The STEs\cite{Seifert2016,Seifert2017} produce THz waves with amplitude comparable to, or even higher than other types of tabletop THz emitters (electro-optic crystals, semiconductor antennas, and air plasma induced by femtosecond laser beam are among the standard methods for producing tabletop THz emitters). Importantly, the efficiency and polarization of THz waves from STEs depends on the magnetization of the FM layer which can be controlled by applying an external magnetic field\cite{Kong2019,Hibberd2019}. However, this reliance on a constant external magnetic field for magnetic saturation of the FM layer for the THz emission limits STEs applications in environments where such fields are undesirable and also impede in downsizing the STEs for spintronic devices. To address this limitation, research on magnetic-field-free STEs is ongoing and holds promise for the next generation of THz technologies. Eliminating the need for external static fields in coherent terahertz emission may open the path for compact, field-free terahertz sources with applications such as THz magneto-optical spectroscopy for field sensitive materials\cite{Wei2023,Shalaby2012}.

THz emission from the spintronic structures without external magnetic field has been previously reported\cite{Kampfrath2013,Schneider2018}. However, more than 50\% reduction in the THz emission amplitude was observed in comparison to the emission in the presence of the external magnetic field. More recently it has been reported that in-plane uniaxial magnetic anisotropy (UMA) in the ferromagnetic layer can be utilized for field-free STEs\cite{Hewett2022,Kueny2023}. These reports proposed a method of oblique angle deposition (OAD) of ferromagnetic layer in CoFeB/Pt bilayer structure and enhancement in UMA by varying the growth conditions of the single CoFeB layer, such as rotating it during the growth process.  However, anisotropy through oblique deposition of the FM layer has limitations in achieving sufficiently high in-plane UMA, as  in this case one is limited to very thin films. Further, the UMA can be deteriorated by annealing due to microstructural modifications or stress relaxation\cite{Dijken2001,Ono1993}. The enhancement in the in-plane UMA induced through oblique deposition of the NM underlayer has been previously explained by the formation of a ripple surface \cite{Scheibler2023,Fukuma2009,McMichael2000}, where the easy axis orientation depends on the details of the ripple structure. Notably, higher UMA leads to high stability against external field disturbances and enables longer-duration field-free STEs, and such STEs could be used for practical field-free THz spintronic emitters. 
STEs utilizing antiferromagnetic (AFM) materials to achieve field-free THz emission were recently reported\cite{Wu2022,Liu2024} with AFM/FM/NM heterostructures. However, the growth of this type of heterostructures,  as well as maintaining a robust exchange bias at the AFM/FM interface are difficult to achieve. This exchange bias arises from the interfacial interaction between the AFM and FM layers, significantly influencing the magnetic properties of the FM layer. However, at room temperature, the formation of a robust exchange bias at the AFM/FM interface presents a significant challenge given by the very thin metallic layers required for efficient THz emission. This limitation is further aggravated by elevated temperatures during the laser-excitation, posing an obstacle to the practical application of AFM-based field-free STEs. Oblique-angle-deposited STEs with enhanced UMA offer a simpler way for field-free THz spintronic emitters.

In this letter, we demonstrate a method to establish in-plane UMA for external magnetic field-free STEs. We introduce ripple structures (through self-shadowing\cite{Barranco2016} and steering effects\cite{Dijken1999}) by obliquely depositing the NM underlayer (W or Ta) instead of the FM layer. We have achieved the external magnetic field-free THz emission from such structures, where the THz emission amplitude remains nearly the same after removing the applied external magnetic field.

We employed the NM 2\,nm / Co$_{40}$Fe$_{40}$B$_{20}$ 2\,nm / Pt 2\,nm (where NM = W or Ta) STEs design known for its high-field THz emission\cite{Seifert2016}. These trilayer structures were deposited with a custom magnetron co-sputtering system with eight 2" sources by Bestec GmbH, Berlin. The target-to-substrate distance was approximately 12\,cm, and the angle of incidence with respect to the sample stage normal was 30$^\circ$. The base pressure was lower than $5\times10{^-}{^8}$\,mbar, and the working gas pressure was $2\times10{^-}{^3}$\,mbar at room temperature. The films were prepared by dc sputtering on double-side polished fused silica substrates with growth rates at 30$^\circ$ incidence angle of 0.116\,nm/s (Ta), 0.064\,nm/s (W), 0.039\,nm/s (CFB), and 0.111\,nm/s (Pt), respectively. No external magnetic field was applied during growth, and no post-annealing was performed. To achieve homogeneous, isotropic samples, the sample stage can be rotated with up to 30\,rpm. In our growth process, the NM underlayers were deposited by varying $\theta$ angles (see Fig.~\ref {fig : Description}(a)) using the standard sample stage, or by equipping it with an additional 30$^\circ$ wedge-shaped holder and aligning it facing the source or away from the source. In all cases, the NM layer was deposited without rotation. This results in three different deposition modes which are summarized in Table~\ref{tab:table1}.

\begin{table}
\caption{\label{tab:table1}Deposition modes of NM (= W or Ta) underlayer}
\begin{ruledtabular}
\begin{tabular}{cccc}
Sample Names & NM at angle\,( $\theta$) & FM/Pt at angle\,($\theta$) & Rotation\,(rpm) \\
\hline
W$30^\circ$ or Ta$30^\circ$ & $30^\circ$ & $30^\circ$ & (FM/Pt) 20\\
W$0^\circ$ or Ta$0^\circ$ & $0^\circ$ & $0^\circ$ & 0\\
W$60^\circ$ or Ta$60^\circ$ & $60^\circ$ & $0^\circ$& 0\\
\end{tabular}
\end{ruledtabular}
\end{table}

\begin{figure*}
\includegraphics{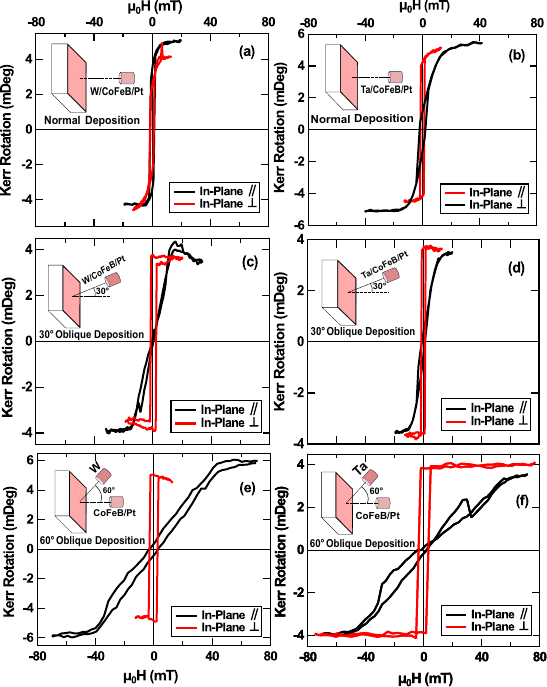}
 \caption{\label{fig : MH} L-MOKE hysteresis loops measurement with the external magnetic field \textit{H} for normal \& oblique deposition of NM (W or Ta) seed layer in NM/CoFeB/Pt structures. (a) \& (b) NM at 0\textdegree{}, (c) \& (d) NM at 30\textdegree{} and (e) \& (f) NM at 60\textdegree .}
\end{figure*}

We have deposited two series of trilayer spintronic structures with either Ta or W as the NM underlayer, where the sample names will be referred to as the name of the underlayer followed by the deposition angle ($\theta$), as summarized in  Table~\ref{tab:table1}. We measured the thickness of the layers using X-ray reflectivity (XRR) with a Rigaku SmartLab (9\,kW rotating anode) system. The deposition times were adjusted to match the 2\,nm layer thickness for every single layer. 

The in-plane UMA was investigated using the longitudinal magneto-optical Kerr effect (L-MOKE) with applied in-plane magnetic fields parallel and perpendicular to the ripple direction. The results are shown in Fig.~\ref{fig : MH}, with the anisotropy increases with increasing angle $\theta$. Samples with oblique deposition show UMA with hard axis (HA) parallel to the plane of incident material flux direction or ripple direction and easy axis (EA) perpendicular to the plane of incident material flux direction or ripple direction. The samples were marked along ripple direction for reference, shown in Fig.~\ref{fig : Description} (c). In particular, we obtain a large anisotropy field with $\mu_0H_a \approx 44$\,mT for $\theta = 60^\circ$, see Fig.~\ref{fig : MH}(e) and (f). This corresponds to a uniaxial anisotropy constant of $K_u = \mu_0H_a M_\mathrm{s}/2 \approx 21.8$\,kJ/m$^3$ with the saturation magnetization $M_s = 990$\,kA/m, as determined by broadband ferromagnetic resonance spectroscopy up to 30\,GHz on an isotropic CFB film of 5\,nm thickness. Both the Ta and W underlayers give similar results for the UMA. A weak UMA in the $\theta = 0^\circ$ samples was traced back to the substrate quality, where the substrates appear to have polishing traces (within their specified roughness) that induce the UMA independent of the sample alignment in the deposition system. This was not observed with higher-quality substrates, such as industrial-grade silicon wafers, which were used for control samples. We thus demonstrated that the UMA can be tuned by changing the incident flux angle of the NM underlayer and a larger UMA can be achieved. We utilized these high-anisotropy spintronic structures for field-free THz emission and present the results in the following.

\begin{figure}
\includegraphics[width = 0.5\textwidth ]{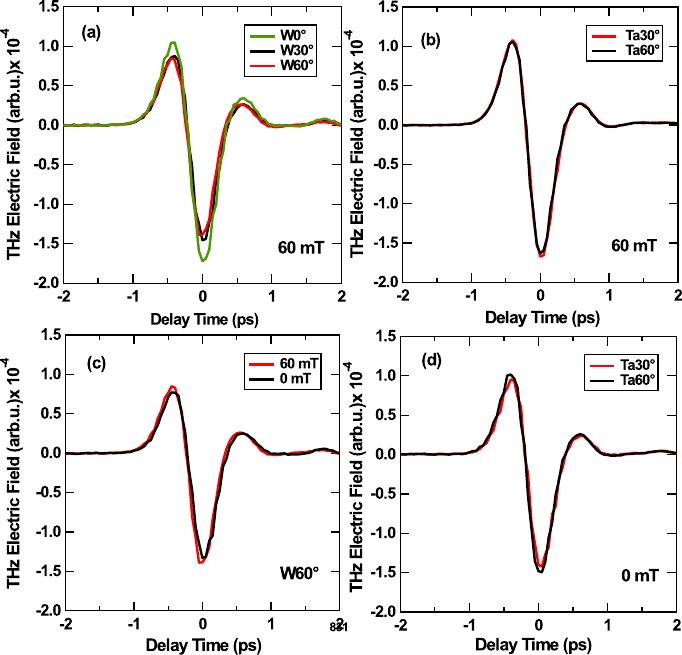}
 \caption{\label{fig : THz}THz signals from the oblique deposited NM (W or Ta) seed layer, in NM/CoFeB/Pt structures with and without the external magnetic field. (a) THz signal emitted from the samples W0\textdegree, W30\textdegree{} and W60\textdegree{} at 60\,mT.  (b) THz signal emitted from the samples Ta30\textdegree{} and Ta60\textdegree{} at 60\,mT. (c) THz signal emitted from the sample W60\textdegree{} at 0 \& 60\,mT. (d) THz signal emitted from the samples Ta30\textdegree{} and Ta60\textdegree{} at 0\,mT.}
\end{figure}

Femtosecond laser pulses used in our THz emission experiments are generated by a Ti:Sapphire laser operating at a 1\,kHz repetition rate with a pulse width of 100\,fs centered at a wavelength of 800\,nm. These pulses were split into pump and probe beams. The pump beam was chopped at 500\,Hz.
The stronger pump pulses with a fluence of $0.40\,\text{m}\text{J}/\text{cm}^2$ (if not stated otherwise) irradiated the sample at normal incidence. The THz emission generated by the excited structures were subsequently detected using electro-optic sampling (EOS)\cite{Wu1995} with a $500\,\mu$m ZnTe crystal detector. The THz emission spectra of the samples are presented in Fig.~\ref{fig : THz} (a) and (b). Here, an external magnetic field of 60\,mT was applied in the plane of the spintronic structure which is enough to saturate the magnetization of ferromagnetic layer and THz emission was recorded.

The results show that the THz emission amplitude from the W0\textdegree{} sample is approximately 17\% higher compared to the W30\textdegree{} and W60\textdegree{} samples. This might be due to the interface roughness introduced with the oblique deposition or due to slight variations of the W layer thickness at different $\theta$, which was not completely compensated with the variation of the deposition times.
Previous studies have shown that the transmission of spin current from the FM layer to the non-magnetic layer depends critically on thickness and surface quality.\cite{Li2019} Furthermore, the external magnetic field was then removed to record the THz emission amplitude in field-free environment for the structures (W60\textdegree, Ta30\textdegree{} and Ta60\textdegree). As shown in the Fig.~\ref{fig : THz} (c) and (d) the THz signal was stable with no significant reduction in THz amplitude at zero field. Experiments over one hour with multiple recorded THz waveforms showed no degradation of the emission amplitude over time without the external field (not shown).

\begin{figure}
\includegraphics[width = 0.49\textwidth]{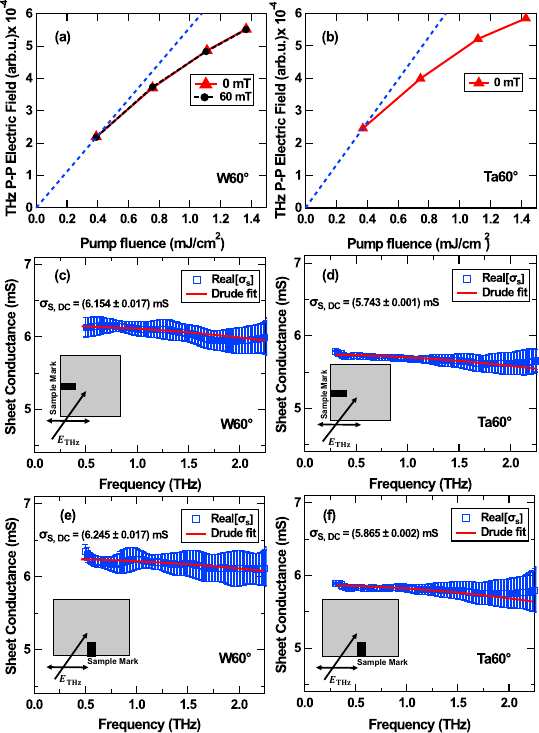}
\caption{\label{fig : THz power} THz emission as a function of pump fluence and a Drude model fit (red solid line) of the real part sheet conductance spectra at room temperature: THz peak-to-peak amplitude (a) W60\textdegree{} at 60\& 0\,mT, and (b) Ta60\textdegree{} at 0\,mT (blue dashed line acts as a guide for the eye), (c) \& (e) sheet conductance spectra for W60\textdegree{} and (d) \& (f) sheet conductance spectra for Ta60\textdegree. Inset: Sample orientation during THz-TDS for measurement of complex sheet conductance spectra, where sample mark represents ripple direction of obliquely deposited NM underlayer.}
\end{figure}

The effect of the pump fluence on the THz field strength was investigated for samples W60\textdegree and Ta60\textdegree{} with and without an external magnetic field. As Fig.~\ref{fig : THz power} shows, the THz amplitude increases with increasing pump fluence.  The excited spin current depends on the electron temperature, which is expected to rise linearly with the pump fluence. Therefore, the simple expectation is a linear dependence of the THz amplitude on the pump fluence. However, a deviation from this linear behaviour is observed in our experiment (see straight lines in Fig.~\ref{fig : THz power} (a) and (b). This is due to the ultrafast demagnetization, only a given amount of spins is available for thermal extraction from the FM layer and leads to the observed saturation\cite{Varela2024,Rouzegar2022}. However, no decrease in the THz amplitude with increasing pump fluence up to $\approx1.4\,\text{m}\text{J}/\text{cm}^2$ is observed, which is in agreement with the previous study\cite{Liu2024}. Importantly, this indicates that there is no damage to the film structure. Also, no hysteresis was observed in the THz emission from the W60\textdegree{} sample when subjected to varying pump fluence ($\approx0.37 - 1.4\,\text{m}\text{J}/\text{cm}^2$) first at 60\,mT and later at 0\,mT, (see Fig.~\ref{fig : THz power}(a)). Here, it is also observed that Ta60\textdegree{} generates 5 to 11\% higher peak-to-peak THz amplitude than W60\textdegree{} under the same pump fluence, which is contrary to the previous findings\cite{Seifert2016}. 

The complex sheet conductance was measured using THz time domain spectroscopy (THz-TDS), and Drude model\cite{Beermann2024} fits of the real part were performed. The imaginary part is effectively zero within experimental error and is not shown in the sheet conductance plots. Drude model fits revealed short scattering times of $\sim$1 to 2\,fs, in agreement with literature on scattering times in metallic thin films\cite{Beermann2024}. The W60\textdegree{} have higher DC sheet conductance than Ta60\textdegree{} (see Fig.~\ref{fig : THz power} (c),(d)) and based on parallel resistivity model\cite{Chen2016} with $\rho_\mathrm{Pt}(2\,\mathrm{nm}) \approx 70\,\mu\Omega\cdot{cm}$\cite{Nguyen2016} and $\rho_\mathrm{CFB}(2\,\mathrm{nm}) \approx 180\,\mu\Omega\cdot{cm}$ it is found that the W layer has lower resistivity $(\approx110\,\mu\Omega\cdot\text{cm})$ than the bulk $\beta$-phase W resistivity (typically around $200\,\mu\Omega\cdot\text{cm})$\cite{Hao2015}.  While tungsten with high resistivity in the $\beta$-phase yields a high spin Hall angle\cite{Chen2018} and strong THz signal, lower-resistivity W will likely result in lower THz emission amplitude, in agreement with our observation. We also observed that the sheet conductivity of both the sample W60\textdegree{} and Ta60\textdegree{} is anisotropic. Specifically, the sheet  contuctance is lower when $E_\mathrm{THz}$ field is perpendicular to the ripple direction (see Fig.~\ref{fig : THz power}(c),(d)) and higher when $E_\mathrm{THz}$ field is parallel to the ripple direction (see Fig.~\ref{fig : THz power}(e),(f)). When $E_\mathrm{THz}$ field is parallel to the ripples, charge carriers experience less scattering as compared to a charge flow transverse to the ripple direction. The observation of the sheet conductance anisotropy is direct evidence for the structural anisotropy in the samples. We note that this result can not be explained by anisotropic magnetoresistance, which would lead to a higher sheet conductance in the configuration with $E_\mathrm{THz}$ perpendicular to the ripple direction.

\begin{figure}
\includegraphics[width = 0.32\textwidth]{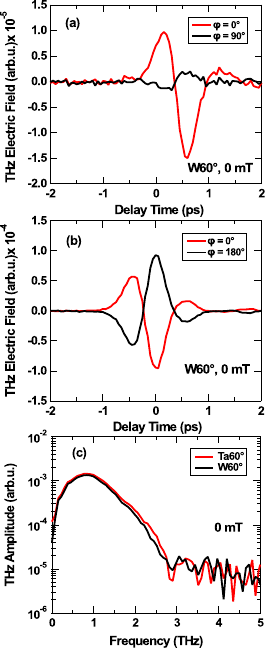}
 \caption{\label{fig : THz phase} The polarization state of the emitted THz waveform and corresponding THz amplitude spectra. (a) THz amplitude measured with polarizer at $\varphi$ = 0\textdegree{} and 90\textdegree{}. (b) THz emission at $\varphi$ = 0\textdegree{} and 180\textdegree{}  (c) THz amplitude spectra by fast fourier transformation.}
\end{figure}

Furthermore, we expect that the THz pulses from the STEs exhibit linear polarization perpendicular to the in-plane magnetization direction\cite{Seifert2016,Li2019,Khusyainov2021}. To determine the polarization state, a polarizer was inserted between the STE and the ZnTe detector. The THz emission was then recorded for the sample W60\textdegree{} with $\varphi$ = 0\textdegree{}, and 90\textdegree{}, (where  $\varphi$ just refers to the rotation of the sample around the laser beam direction). This rotation resulted in a complete suppression of the THz signal, indicating a well-defined polarization direction for the emitted THz pulse, see Fig.~\ref{fig : THz phase} (a). The polarization of the STEs can be controlled by both the applied magnetic field and the azimuthal rotation of the spintronic structure about the surface normal. Phase control through azimuthal rotation in zero external magnetic field is demonstrated in Fig.~\ref{fig : THz phase}(b) by rotating the sample 180\textdegree{} relative to its initial orientation. However, in presence of the external magnetic field THz emission is, as expected, invariant of azimuthal rotation. In the THz amplitude spectra  (Fig.~\ref{fig : THz phase}(c)), the spectral width is limited by both the $500\,\mu$m thick ZnTe crystal detector and the 100\,fs laser pulse duration used in this investigation. Shorter laser pulse durations and a relatively thinner ZnTe crystal detector could be used during the experiment for higher detected spectral bandwidth.

In summary, we have demonstrated in-plane uniaxial magnetic anisotropy in the ferromagnetic layer by utilizing an obliquely deposited non-magnetic underlayer in trilayer spintronic THz emitter structures. Notably, all structures exhibited anisotropies with the easy axis perpendicular to the ripple direction. The spintronic structures with high anisotropies were successfully employed to generate terahertz emission without the need for an external magnetic field. The emitted THz radiation exhibits linear polarization, and its polarization can be controlled by both varying the direction of an applied external magnetic field (when present) and by the azimuthal rotation of the sample in the absence of an external magnetic field. Operating spintronic THz emitters in a field-free environment not only eliminates a disadvantage with respect to electro-optic crystals and semiconductor antenna tabletop THz sources but also opens doors for integration, high-density multi-emitter systems, practical applications in communication, and security imaging\cite{Li2023}.

\section*{ACKNOWLEDGMENT}
\vspace{-\baselineskip}
The Darmstadt group acknowledges the financial support by the Deutsche Forschungsgemeinschaft under Project No. 513154775 and No. 518575758, and by the DFG Major Research Instrumentation programme Project No. 511340083 and Project No. 468939474. The group in Bielefeld acknowledges the financial support from the European Union’s Horizon 2020 research and innovation program (Grant Agreement No.964735 EXTREME-IR), Deutsche Forschungsgemeinschaft (DFG) within Project No. 468501411-SPP2314 INTEGRATECH and Project No. 518575758 HIGHSPINTERA, Bundesministerium für Bildung und Forschung (BMBF) within Project No. 05K2022 PBA Tera-EXPOSE, and Bielefelder Nachwuchsfond. We, gratefully acknowledge Prof. Lambert Alff at Technical University Darmstadt for providing us with access to his  laboratory's X-ray diffractometer, which was instrumental in carrying out this research work.

\section*{author declarations}
\vspace{-\baselineskip}

\subsection*{Conflict of Interest}
\vspace{-\baselineskip}
The authors have no conflicts to disclose.

\section*{Data Availability}
\vspace{-\baselineskip}
The data that support the findings of this study are available from the corresponding author upon reasonable request.
\section*{References}

\bibliography{library}

\end{document}